\documentclass[12pt]{article}
\usepackage[pctex32]{graphics}
\begin{document}
~
\vspace{0.5cm}
\begin{center} {\Large \bf  Semiclassical Strings in  Electric and Magnetic Fields Deformed $AdS_5 \times S^5$ Spacetimes}
                                                  
\vspace{1cm}

                      Wung-Hong Huang\\
                       Department of Physics\\
                       National Cheng Kung University\\
                       Tainan, Taiwan\\

\end{center}
\vspace{1cm}
\begin{center} {\large \bf  Abstract} \end{center}
We first apply the transformation of mixing azimuthal and internal coordinate or  mixing time and internal coordinate to the 11D M-theory with a stack N M2-branes to find  the spacetime of a stack of  N D2-branes with magnetic or electric flux  in 10 D IIA string theory, after the Kaluza-Klein reduction.  We then perform the T duality to the spacetime to find the background of a stack of  N D3-branes with magnetic or electric flux.  In the near-horizon limit the background becomes the magnetic or electric field deformed $AdS_5 \times S^5$.  We adopt an ansatz  to find the classical string solution which is rotating in the deformed $S^5$ with three angular momenta in the three rotation planes.  The relations between the classical string energy and its angular momenta are found and results show that the external magnetic and electric fluxes will increase the string energy.  Therefore, from the AdS/CFT point of view, the corrections of the anomalous dimensions of operators in the dual SYM theory will be positive.
We also investigate the small fluctuations in these solutions and discuss the effects of magnetic and electric fields on the  stability of these classical rotating string solutions.  Finally, we find the possible solutions of string pulsating on the deformed spacetimes and show that the corrections to the anomalous dimensions of operators in the dual SYM theory are non-negative.

\vspace{2cm}
\begin{flushleft}
E-mail:  whhwung@mail.ncku.edu.tw\\
\end{flushleft}
\newpage
\section {Introduction}
It is known that the AdS/CFT correspondence plays important role in studying the gauge theories at strong coupling [1,2].  Extending it to less supersymmetric cases [3-16] may allow us to find  simple  string-theoretic  descriptions  of  various dynamical aspects of gauge theories, from high-energy scattering to confinement in phenomena.  The generalization of AdS/CFT duality to the non-BPS string mode sector can be guided by semiclassical considerations as suggested  in [3-6].   In particular, the investigations of [5,6] had found that using the novel multi-spin string states one can  carry out  the  precise test of the  AdS/CFT duality in a  non-BPS sector by comparing the  ${\lambda \over J^2 }\ll 1$  expansion of the {\it classical} string energy  with  the corresponding {\it quantum}  anomalous dimensions in  perturbative SYM theory [7-9]. 

In recent Lunin and Maldacena [10] had demonstrated that certain deformation of the $AdS_{5}\times S^{5}$ background corresponds to a $\beta$-deformation of $N=4$ SYM gauge theory in which the supersymmetry being broken was studied by Leigh and Strassler [11].  Since then there are many studies of AdS/CFT correspondence in the $\beta$-deformation Lunin-Maldacena background [12-15].

    In a previous publication [16] we considered the magnetic-flux deformed $AdS_5 \times S^4$ and $AdS_4 \times S^5$ background which are obtained by  performing  the dimensional reduction of the 10D spacetime of the N D3-branes through the transformation of mixing azimuthal and internal coordinate, the  ``twist'' identification of a circle [17,18].  The new backgrounds may be regarded as those with Melvin magnetic flux [19].  As the supersymmetry in the deformed $AdS_m\times S^n$ backgrounds has been broken by the presence of magnetic field they offer the spacetimes to study the AdS/CFT correspondence with less supersymmetry. 

In this paper we will apply the transformation of mixing azimuthal and internal coordinate [17,18 ] or mixing time and internal coordinate [20,21]  to the 11D M-theory with a stack $N$ M2 branes [22] and then use T duality [23,24] to find the spacetime of a stack of  N D3-branes with external magnetic or electric field.  In the near-horizon limit the background becomes the magnetic or electric field  deformed $AdS_5 \times  S^5$.  As the isometry groups of  $AdS_5 \times S^5$ are $SO(2,4) \times SO(6)$ which are exactly the conformal group and R-symmetry of $N=4$ supper Yang-Mills theory the background studied in this paper will be more phenomenally interesting.

We will first use the method in [5,6] to find the classical rotating string solutions which are rotating in deformed $S^5$ with three equal angular momenta $J$ in the three rotation planes.   The relations between the classical string energy and its angular momenta are found.   The results show that the external magnetic and electric fluxes will increase the classical string energy.  We also investigate the small fluctuations in these solution and discuss the effects of  the magnetic and electric fields on the stability of these classical rotating string solutions.  As the corrections to the classical string energy, from the AdS/CFT point of view, give the anomalous dimensions of operators in the dual SYM theory our results therefore are of primary interest.  We next use the method in [25,26] to find the classical pulsating string solutions and show that the corrections to the anomalous dimensions of operators in the dual SYM theory are non-negative.

In section II we study the case of classical rotating string in the electric field deformed $AdS_5 \times  S^5$ and in section III the case in the magnetic field deformed $AdS_5 \times  S^5$.  In section IV we study the string pulsating on the deformed spacetimes. In section V we discuss our results.

\section{Rotating String in Electric Field Deformed $AdS_5 \times S^5$}
\subsection{Electric Field Deformed $AdS_5 \times S^5$}
We now  apply the transformation of mixing time and internal coordinate [20,21] to the 11D M-theory with a stack $N$ M2-branes [22] and then use the T duality [23,24] to find the spacetime of a stack of  N D3- branes with electric flux. 

The full $N$ M2-branes solution is given by [22]
$$ds^2_{11}=H^{-2\over3}\left(-dt^2+dx_1^2+dx_2^2\right)+H^{1\over3}\left(dx_3^2 + d\rho^2+\rho^2 \Omega_5^2+dx_{11}^2\right),$$
$$ ~~~~~~~~~~d\Omega_5^2 \equiv d\gamma^2+cos^2\gamma d\varphi_1^2+sin^2\gamma(d\psi^2 + cos^2\psi\ d\varphi_2^2+ sin^2\psi\ d\varphi_3^2), \eqno{(2.1)}$$
$$A^{(3)} = H^{-1} dt\wedge dx_1\wedge dx_2.\eqno{(2.2)}$$
$H$ is the harmonic function defined by 
$$ H = 1+ {R\over r^{D-p-3}}, ~~~~~~~r^2\equiv x_3^2+ \rho^2+x_{11}^2 , ~~~~R \equiv {16\pi G_D\,T_p \,N\over D-p-3 },\eqno{(2.3)}$$
in which  $G_D$ is the D-dimensional Newton's constant and $T_p$ the p-brane tension. In the case of (2.1), $D=11$ and $p=2$.

 Now we transform the time $t$ by mixing it with the compactified coordinate $x_{11}$ in the following substituting [20,21]
$$ t  \rightarrow t - E x_{11}.\eqno{(2.4)}$$ 
Using above substitution the line element (2.1)  becomes
$$ds^2_{11}={-H^{-2\over3} dt^2\over 1- E^2 H^{-1}} + H^{-2\over3}\left(dx_1^2+dx_2^2\right)+H^{1\over3}\left(dx_3^2 + d\rho^2+\rho^2 d\Omega_5^2  \right)$$
$$~~~~~~ +\left(H^{1\over3}- E^2 H^{-2\over3}\right) \left(dx_{11} +{EH^{-1}\over 1- E^2 H^{-1}}dt \right)^2, \eqno{(2.5)}$$
Using the relation between the 11D M-theory metric and string frame metric, dilaton field and 1-form potential
$$ds_{11}^2= e^{-2\phi/3}ds_{10}^2+  e^{4\phi/3} (dx_{11}+2 A_\mu dx^\mu )^2 , \eqno{(2.6)} $$
the 10D IIA background is described by
$$ds_{10}^2 ={-H^{-1\over2} \over \sqrt {1- E^2 H^{-1}}}\,dt^2 + \sqrt {1- E^2 H^{-1}}\left[H^{-1\over2}\left(dx_1^2+dx_2^2\right)+H^{1\over2}\left(dx_3^2 + d\rho^2+\rho^2 d\Omega_5^2 \right)\right],\eqno{(2.7)}$$ 
$$e^{4\Phi\over 3}= H^{1\over 2}\left(1-E^2 H^{-1}\right), ~~~~~ A_{t} ={EH^{-1}\over 1-E^2 H^{-1}} .\hspace{2cm}\eqno{(2.8)}$$
In this decomposition into ten-dimensional fields which do not depend on the $x_{11}$, the ten-dimensional Lagrangian density becomes 
$$ L = R- 2 (\nabla \varphi)^2 - e^{2\sqrt 3 \varphi}~ F_{\mu\nu}F^{\mu\nu},\eqno{(2.9)}$$
and from (2.8) we see that the parameter $E$ represents the magnitude of the external electric field.  In the case of  $E=0$  the above spacetime becomes the well-known geometry of a stack of N D2-branes.  Thus, the background describes the spacetime of a stack of  N D2-branes with electric flux.

  To find the spacetime of a stack of N D3-branes we now perform the T-duality transformation [23,24] on the coordinate $x_{3}$.  After substituting the metric and dilation field by 
$$g_{x_3 x_3}\rightarrow {1\over g_{x_3 x_3}}, ~~~~~~e^{\Phi}\rightarrow {e^{\Phi}\over \sqrt{g_{x_3 x_3}}},\eqno{(2.10)}$$
the metric describing a stack of N D3-branes with electric flux becomes 
$$ds_{10}^2 ={-H^{-1\over2} \over \sqrt {1- E^2 H^{-1}}}\,\left(dt^2-dx_3^2\right)+ \sqrt {1- E^2 H^{-1}}\left[H^{-1\over2}\left(dx_1^2+dx_2^2\right)+H^{1\over2}\left(d\rho^2+\rho^2 d\Omega_5^2 \right)\right]. \eqno{(2.11)}$$ 
Far from the sources, namely for $\rho \gg R$, above metric approaches Minkowski space, since $H \sim 1$. In the ``near-horizon'' limit, namely the region $\rho \ll R$,  we can approximate $H \sim \frac{R}{r^4}$, and the line element (2.11) becomes
$$ds_{10}^2 = -{\rho^2\over R \sqrt{1-{E^2\rho^4\over R}}}\left(dt^2 - dx_3^2\right) +\sqrt{1-{E^2\rho^4\over R}}\left[{\rho^2\over R} \left(dx_1^2+dx_2^2\right) + {R\over \rho^2} \left(d\rho^2+\rho^2 d\Omega_5^2\right)\right]. \eqno{(2.12)}$$ 
In the case of  $E=0$  the above spacetime becomes the well-known geometry of $AdS_5\times S^5$. Thus, the background describes the electric field deformed $AdS_5\times S^5$. 

\subsection{Rotating String Solution in Electric Field Deformed $S^5$}
We will search the string solutions which is fixed on the spatial coordinates in deformed $AdS_5$ locating at  $x_1=x_2=x_3=0$ in the electric field deformed spacetime (2.12) with a fixed value of $\rho$.  The line element becomes
$$ds_6^2 = -{1\over\sqrt{1-E^2}}\,dt^2 +\sqrt{1-E^2}\left[d\gamma^2+cos^2\gamma d\varphi_1^2+sin^2\gamma(d\psi^2 + cos^2\psi\ d\varphi_2^2+ sin^2\psi\ d\varphi_3^2)\right].\eqno{(2.13)}$$ 
In the above equation, as we merely want to see the effect of electric filed on the string solution we let $R=\rho =1$ for convenience, which means that we have used the scale  $ {E^2 \rho^4 \over R} \rightarrow  E^2$. 

  The string action could be written in the conformal gauge in terms of the independent  global coordinates $x^m$  
$$ I= - { 1 \over 4 \pi} \int d^2 \xi \  G_{mn}(x) \partial_a x^m \partial^a  x^n, \eqno{(2.14)}$$
in which $\xi^a=(\tau,\sigma)$ and we let $\alpha' = 1$ for convenience.  In the conformal gauge $\sqrt {-g} g^{ab} = \eta^{ab}= $diag(-1,1), the equations of motion following from the action should be
supplemented by the conformal gauge  constraints
$$ G_{mn}(x) ( \dot  x^m \dot  x^n +  x'^m   x'^n) =0 , \eqno{(2.15a)}$$
$$ G_{mn}(x) \dot x^m   x'^n =0.  \eqno{(2.15b)}$$
Following the method of Frolov and Tseytlin [5,6] we now adopt  the ansatz 
$$ t =\kappa \tau,~~~ \gamma = \gamma(\sigma),~~~ \psi = \psi(\sigma),~~~\varphi_1= \nu \tau,~~~\varphi_2 = \varphi_3= \omega \tau.  \eqno{(2.16)}$$
to find the rotating string solution. 

Substituting the ansatz (2.16) into metric form (2.13) the associated Lagrangian  is 
$$L = - {1\over 4 \pi}\left[{1\over \sqrt{1-E^2}}\kappa^2 - \sqrt{1-E^2}\left( \nu^2 cos^2\gamma(\sigma) + \omega^2 sin^2\gamma(\sigma) - \gamma(\sigma)'^2  - \psi(\sigma)'^2 sin^2\gamma(\sigma) \right) \right].\eqno{(2.17)}$$
As the deformation we used does not change the properties of the  translational isometries of  coordinates $t$, $\varphi_1$, $\varphi_2$ and $\varphi_3$, there are the corresponding four integrals of motion:
$$ {\cal E}= P_{t}= \int^{2\pi}_0 {d\sigma\over 2\pi}{1\over  {\sqrt{1-E^2}}}\,\partial _0 t \,,  \eqno{(2.18)}$$
which is the energy of the solution, and  
$$ J_1= P_{\varphi_1}= \int^{2\pi}_0{d\sigma\over 2\pi} {\sqrt{1-E^2}}~cos^2\gamma(\sigma)\,\partial _0\varphi_1 , \hspace{1.7cm} \eqno{(2.19)}$$
$$ J_2= P_{\varphi_2}=  \int^{2\pi}_0{d\sigma\over 2\pi}{\sqrt{1-E^2}}~sin^2\gamma(\sigma)\,\,cos^2\psi(\sigma)~\partial _0\varphi_2 ,  \eqno{(2.20)}$$
$$ J_3= P_{\varphi_3}= \int^{2\pi}_0 {d\sigma\over 2\pi}{\sqrt{1-E^2}}~sin^2\gamma(\sigma)\,\,sin^2\psi(\sigma)~\, \partial _0\varphi_3,  \eqno{(2.21)}$$
which are the angular momenta of the rotating string in the electric field deformed $S^5$ space. 

 To find the values of energy and angular momenta we must know the function of $\gamma(\sigma)$ and $\psi(\sigma)$, and have  relations between $\kappa$, $\nu$ and $\omega$.  This can be obtained by solving the equations of  $\gamma(\sigma)$ and $\psi(\sigma)$ associated to the Lagrangian (2.17), and imposing the conformal gauge constraints of (2.15).  The field equation of $\psi(\sigma)$ is
$$0= \left( \psi'(\sigma)~sin^2\gamma(\sigma) \right)',\eqno{(2.22)}$$
which could be easily solved by setting
$$ \psi(\sigma) = n \sigma,~~~~\gamma(\sigma) =\gamma_0, \eqno{(2.23)}$$
which are the same as those in the undeformed space [5].  In this case the field equation of $\gamma(\sigma)$ reduces to a simple relation 
$$ \nu^2 = \omega^2-n^2. \eqno{(2.24)}$$
which is also the same as that in the undeformed space [5].  While the conformal gauge constraints (2.15b) is automatically satisfied the another conformal gauge constraints of (2.15a) implies
$$\kappa^2 =\left(1-E^2\right)\left( \nu^2+ 2 n^2sin\gamma_0^2\right).\eqno{(2.25)}$$
Using the above relations the energy and angular momenta of the string solution have the simple forms
$${\cal E}= \sqrt{\nu^2+ 2n^2sin^2 \gamma_0}, ~~~ J_1= \sqrt{1-E^2} \,\nu\,cos^2\gamma_0,~~~~J\equiv J_2=J_3= {1\over2}\sqrt{1-E^2} \,\sqrt{\nu^2+n^2}\,sin^2\gamma_0,\eqno{(2.26)}$$
respectively.  We now use the above results to analyze two cases.
\\

1.   $\nu =0$:  In this case, using (2.26) we see that  ${\cal E}= \sqrt{2}\,n \,sin\gamma_0$, $J_1=0$ and $J={1\over2}\sqrt{1-E^2}\,n\,sin^2\gamma_0$, thus 
$${\cal E}= {2\over\left(1-E^2\right)^{1\over 4}}\sqrt {nJ} > 2\sqrt{ nJ}.\eqno{(2.27)}$$
Therefore the external electric fluxes will increase the string energy.  To analyze the stability of  the above solution we recall that, without the $E$ field the problems of the stability of the rotating string in the $S^5$ had been investigated in detail  in [5] by Frolov and Tseytlin.  The result of appendix A.2 in [5] had shown that the rotating string is stable only if 
$$0 \le \kappa^2 \le {3\over 2}.\eqno{(2.28)}$$
We claim that the above criterion could still be used in the electric field deformed $S^5$ background.  This is because that the Lagrangian associated to the metric eq.(2.13) used to investigate the fluctuation field in the deformed case is equal to that used in the undeformed case, up to an overall constant value  ``$\sqrt{1- E^2}$'', after rescaling the time by $t \rightarrow {t\over \sqrt{1-E^2}}$.   Therefore, the criterion (2.28) would not be changed in the case with electric field deformation.  Now, from (2.25), (2.26) and (2.27) we know that $\kappa^2 = \left(1-E^2\right) {\cal E}^2= {4\,\left(1-E^2\right)^{1\over 2}}n J$.  Thus the criterion  (2.28) becomes
$$0 <  J < J_c \equiv{3\over 8n \left(1-E^2\right)^{1\over 2}}.\eqno{(2.29)}$$
As $  J_c > {3\over 8n}$ the electric fields therefore have inclinations to improve the  stability of these classical rotating string solutions.  
\\

2.   $\nu \gg n $:  In this case, using (2.26) we see that  $J_1+ 2J \approx \sqrt{1-E^2} \,\nu$ and $sin^2\gamma_0 \approx {2J\over \sqrt{1-E^2}\,\nu} = {2J\over J_1+2J}$, thus 
$${\cal E}\approx {J_1+2J\over\sqrt{1-E^2}}.\eqno{(2.30)}$$
Therefore the external electric fluxes will increase the string energy.    To analyze the stability of  the above solution we recall that, without the $E$ field the problems of the stability of the above rotating string in the $S^5$ had been investigated in detail  in [6] by Frolov and Tseytlin.  The result of eq.(2.36) in [6] had shown that the rotating string is stable only if 
$$0 \le sin^2\gamma_0 \le {3\over 4}.\eqno{(2.31)}$$
Following the discussion in the previous case we see that the above criterion could still be used in the electric field deformed $S^5$ background.  In the case of  $\nu \gg 1 $ we can substitute the relation $sin^2\gamma_0 \approx {2J\over J_1+ 2J}$ into the criterion (2.23) and find that
$$0 < {2J\over J_1+ 2J} \le {3\over 4} ~~~~\Rightarrow~~~{J\over J_1} \le {3\over2}.\eqno{(2.32)}$$
This means that the electric fields does not change stability of these classical rotating string solutions. Let us make two comments to conclude this section.

  1. As the correction to the rotating classical string energy is positive then, from the AdS/CFT point of view, the correction of the anomalous dimensions of  operators in the dual SYM theory will be positive.

  2.  In principle, there are the dilaton and anti-symmetric fields terms in the string Lagrangian.  However, as the induced metric of the rotating string solution found in this section is flat and there is not NS-NS $B_2$ field in our background the action (2.14) is a proper one to be used. The property also reveals in the section III.


\section{Rotating String in Magnetic Field Deformed $AdS_5 \times S^5$}
\subsection{Magnetic Field Deformed $AdS_5 \times S^5$}
We now  apply the transformation of mixing azimuthal and internal coordinate [17,18 ]  to the 11D M-theory with a stack $N$ M2-branes [22] and then use the T duality [23,24] to find the spacetime of a stack of  N D3- branes with magnetic flux. 

Using the full N M2-branes metric described in (2.1) we can transform the angle $\varphi_1$ by mixing it with the compactified coordinate $x_{11}$ in the following substituting
$$\varphi_1 \rightarrow \varphi_1 + B x_{11}.\eqno{(3.1)}$$ 
Using the above substitution the line element (2.1)  becomes
$$ds_{11}^2 = H^{-2\over3} \left(- dt^2+ dx_1^2+dx_2^2 \right) + H^{1\over3}\left(dx_3^2+d\rho^2 + \rho^2\left( d\gamma^2 + {cos^2\gamma\ d\varphi_1^2\over 1+ B^2~\rho^2 cos^2\gamma}\ d\varphi_1^2 \right.\right.$$
$$\left.\left.+ sin^2\gamma d\Omega^2_3 \right) \right) + H^{1\over3}\left(1+ B^2\rho^2 cos^2\gamma \right)\left(dx_{11} + {B\rho^2cos^2\gamma\ d\varphi_1^2\over 1+ B^2 ~\rho^2 cos^2\gamma}\right)^2\, ,$$ 
$$d\Omega^2_3 \equiv d\psi^2 + cos^2\psi\ d\varphi_2^2+ sin^2\psi\ d\varphi_3^2. \eqno{(3.2)}$$
Using the relation between the 11D M-theory metric and string frame metric, dilaton field and 1-form potential described in (2.6) the 10D IIA background is then described by
$$ds_{10}^2 =H^{-1\over2}\sqrt{1+ B^2\rho^2 cos^2\gamma} \left(- dt^2+ dx_1^2+dx_2^2 \right) + H^{1\over2}\sqrt{1+ B^2\rho^2 cos^2\gamma}\left(dx_3^2+d\rho^2\right.$$
$$ ~~~~~~\left.+ \rho^2\left( d\gamma^2 + {cos^2\gamma\ d\varphi_1^2\over 1+ B^2~\rho^2 cos^2\gamma}\ d\varphi_1^2 + sin^2\gamma\ d\Omega_3^2 \right) \right).\eqno{(3.3)}$$ 
$$e^{\Phi} = \sqrt{1+ B^2\rho^2 cos^2\gamma}~H^{1\over4}, ~~~~~ A_{\phi_1} ={B\rho^2~cos^2\gamma\over 2\left(1+ B^2\rho^2 cos^2\gamma\right)}.\hspace{2cm}\eqno{(3.4)}$$
In this decomposition into ten-dimensional fields which do not depend on the $x_{11}$, the ten-dimensional Lagrangian density will be described by (2.9) 
and the parameter $B$ is the magnetic field defined by $B^2=\frac{1}{2}F_{\mu\nu}F^{\mu\nu}|_{\rho=0}$.   In the case of  $B=0$  the above spacetime becomes the well-known geometry of a stack of N D2-branes.  Thus, the background describes the spacetime of a stack of  N D2-branes with magnetic flux.

  To find the spacetime of a stack of N D3-branes we now perform the T-duality transformation [23,24] on the coordinate $x_{3}$.  Using the substitution (2.10) 
the background describing a stack of N D3-branes with magnetic flux therefore becomes 
$$ds_{10}^2 =H^{-1\over2}\left[\sqrt{1+ B^2\rho^2 cos^2\gamma} \left(- dt^2+ dx_1^2+dx_2^2 \right) + {1\over\sqrt{1+ B^2\rho^2 cos^2\gamma}}dx_3^2 \right]\hspace{2cm}$$
$$+ H^{1\over2}\sqrt{1+ B^2\rho^2 cos^2\gamma}\left(d\rho^2 + \rho^2\left( d\gamma^2 + {cos^2\gamma\ d\varphi_1^2\over 1+ B^2~\rho^2 cos^2\gamma}\ d\varphi_1^2 + sin^2\gamma\ d\Omega_3^2 \right) \right).\eqno{(3.5)}$$ 
$$e^{\Phi} = \sqrt{1+ B^2\rho^2 cos^2\gamma}, ~~~~~ A_{\phi_1} ={B\rho^2~cos^2\gamma\over 2\left(1+ B^2\rho^2 cos^2\gamma\right)}.\hspace{2cm}\eqno{(3.6)}$$
Far from the sources, namely for $\rho \gg R$, above metric approaches Minkowski space, since $H \sim 1$. In the ``near-horizon'' limit, namely the region $\rho \ll R$,  we can approximate $H \sim \frac{R}{r^4}$, and the line element (3.5) becomes
$$ds_{10}^2 = R \sqrt{1+B^2R^2Z^{-2}cos^2\gamma}\left[{1\over Z^2}(- dt^2+ dx_1^2+dx_2^2 +{1\over {1+B^2R^2Z^{-2}}}dx_3^2+dZ^2) +\right.\hspace{2cm} $$
$$\left.\hspace{6cm} \left(d\gamma^2 + {cos^2\gamma\ d\varphi_1^2\over 1+ B^2R^2Z^{-2}cos^2\gamma} + sin^2\gamma\ d\Omega_3^2 \right)\right],\eqno{(3.7)}$$ 
in which we define $Z\equiv R^2/\rho$.   In the case of  $B=0$  the above spacetime becomes the well-known geometry of $AdS_5\times S^5$. Thus, the background describes the magnetic-deformed $AdS_5\times S^5$. 

\subsection{Rotating String Solution in Magnetic Field Deformed $S^5$}
We will search the string solutions which is fixed on the spatial coordinates in deformed $AdS_5$ locating at  $x_1=x_2=x_3=0$ in the magnetic field deformed spacetime (3.7) with a fixed value of $Z$.  The line element becomes
$$ds_{6}^2 = \sqrt{1+B^2 cos^2\gamma}\left[- dt^2 + d\gamma^2 + {cos^2\gamma d\varphi_1^2\over 1+ B^2 cos^2\gamma} + sin^2\gamma\ d\Omega_3^2 \right],\eqno{(3.8)}$$ 
in which, as we merely want to see the effect of magnetic Melvin filed on the string solution, we let $R=\rho=1$ for convenience.   This means we have used the scale  $ {B^2 R^2 Z^{-2}} \rightarrow  B^2$. 

  To proceed, we use the classical rotating string solution ansatz  (2.16) to find the associated Lagrangian  
$$L = - { 1 \over 4 \pi }\sqrt{1+ B^2\,cos\gamma(\sigma)}\left[-\kappa^2 + {\nu^2 cos^2\gamma(\sigma)\over 1+ B^2cos^2\gamma(\sigma)}\, + \omega^2sin^2\gamma(\sigma) - \gamma(\sigma)'^2  - \psi(\sigma)'^2 sin^2\gamma(\sigma)  \right],\eqno{(3.9)}$$
Using the  Lagrangian we can find that the field equation of $\psi(\sigma)$ is
$$0= \left( \psi'(\sigma)~sin^2\gamma(\sigma) \right)', \eqno{(3.10)}$$
which could be easily solved by setting
$$ \psi(\sigma) = n \sigma,~~~~\gamma(\sigma) =\gamma_0, \eqno{(3.11)}$$
which are the same as those in the undeformed space [5].   Using this relation the field equation of  $\gamma(\sigma)$
$$0= {\partial\over\partial\, \gamma_0}\left(\sqrt{1+ B^2\,cos^2\gamma_0}\left[-\kappa^2 + {\nu^2 cos^2\gamma_0\over 1+ B^2cos^2\gamma_0}\, + \omega^2sin^2\gamma_0- n^2 sin^2\gamma_0 \right]\right) , \eqno{(3.12)}$$
implies a simple relation
$$\kappa^2 = \left[ {2\over B^2} - {cos^2\gamma_0\over 1+ B^2cos^2\gamma_0}\right] \nu^2~ - (\omega^2-n^2)\left[{2\over B^2}+ 3cos^2\gamma_0-1\right]. \eqno{(3.13)}$$
In the case of $B^2 \rightarrow 0$ above equation reduces to the relation $\omega^2 = \nu^2+n^2$ as that in [5]. To proceed, we see that while the conformal gauge constraints (2.15b) is automatically satisfied the another conformal gauge constraints of (2.15a) implies that 
$$\kappa^2 =  {cos^2\gamma_0\over 1+ B^2cos^2\gamma_0} \nu^2+ (n^2+ \omega^2) ~sin\gamma_0^2.\eqno{(3.14)}$$
From (3.13) and (3.14) we can find that
$$\omega^2 = {\nu^2\over (1+B^2cos^2\gamma_0)^2}+ {n^2 \left(1+B^2(2cos^2\gamma_0-1)\right)\over 1+B^2cos^2\gamma_0}.\eqno{(3.15)}$$
Now we can follow the definition in (2.18)-(2.21) to evaluate the energy and angular momenta of the rotating string.  The formula are 
$${\cal E}= \kappa \,\sqrt{1+B^2cos^2\gamma_0}= \sqrt{1+B^2cos^2\gamma_0}\hspace{9cm}$$
$$~~~~~\times\sqrt{{cos^2\gamma_0\over 1+ B^2cos^2\gamma_0} \nu^2+\left({\nu^2\over (1+B^2cos^2\gamma_0)^2}+ {n^2 \left(2+B^2(3cos^2\gamma_0-1)\right)\over 1+B^2cos^2\gamma_0}\right)sin^2\gamma_0}~,\eqno{(3.16)}$$
$$J_1 = {\nu\,cos^2\gamma_0\over \sqrt{1+ B^2cos^2\gamma_0}},\hspace{13.5cm}\eqno{(3.17)}$$
$$J\equiv J_2=J_3 = {\omega\over2}\sqrt{1+ B^2cos^2\gamma_0}\, sin^2\gamma_0 =  {1\over2}\sqrt{1+ B^2cos^2\gamma_0}\, sin^2\gamma_0\,\hspace{5cm}$$
$$\hspace{5cm}~~~~\times \sqrt{{\nu^2\over (1+B^2cos^2\gamma_0)^2}+ {n^2 \left(1+B^2(2cos^2\gamma_0-1)\right)\over 1+B^2cos^2\gamma_0}} .\eqno{(3.18)}$$
From (3.17) and (3.18) we can, in principle, express $\nu$ and $\gamma_0$ as the functions of $J_1$ and $J$.  Then,  substituting the functions into (3.16) we can express the energy ${\cal E}$ as the functions of $J_1$ and $J$.  However, as the algebra in there is too complex we will first consider the case of $\nu = 0$ under a small magnetic flux. 
\\

1.  $\nu = 0$:  In this case, using (3.18) we can express  $cos^2\gamma_0$ as a function of $ J$,  i.e.
$$cos^2\gamma_0 = (1-{2J\over n})+{J\over n}\left(1- {4J\over n}\right) B^2+O(B^4).\eqno{(3.19)}$$
Substituting this relation into (3.16) we obtain a simple form of the string energy 
$${\cal E} = 2\sqrt {nJ}+ {1\over 2} \sqrt {nJ}\,\left(1- {2J\over n}\right)B^2 + O(B^4).\eqno{(3.20)}$$
The second term is the corrected energy raised from the Melvin magnetic flux which deforms the $S^5$.  As a rotating string with $ {1\over2} \le{J\over n}$ will be a unstable solution, which is investigated in the following, the above result shows that the stable rotating string has a positive corrected energy.
 
   To analyze the stability of the above rotating string solution we can also use the criterion (2.28), $0 \le \kappa^2 \le {3\over 2}$, in the case of weak magnetic flux.   The reason is that, from (3.8) we see that the magnetic flux $B^2$ only appears in the combination form ``${1+ B^2cos^2\gamma(\sigma)}$''.  Thus, during considering the fluctuation of the field $\gamma$ we shall replace $\gamma \rightarrow \gamma_0 +\tilde\gamma(t,\sigma)$ in the original Lagrangian.  Then the combination form could be approximated by $1+ B^2cos^2(\gamma_0+\tilde\gamma(t,\sigma)) \approx 1+ B^2cos^2\gamma_0,$ in the case with  a small value of $B^2$.  Thus, the Lagrangian used to investigate the fluctuation field $\gamma$ in the deformed case is equal to that used in the undeformed case, up to an overall constant value  ``$\sqrt{1+ B^2cos^2\gamma_0}$'', after rescaling the field by $\varphi_1 \rightarrow \left(1+ B^2cos^2\gamma_0\right)^{-1/2}\varphi_1$.   Therefore, the criterion (2.28) would not be changed under a small magnetic-flux deformation.    Now, as found in (3.19) we have expressed  $cos^2\gamma_0$ as a function of $ J$, we can now substitute this relations into (3.14) and (3.15) and, therefore can express $\kappa$ as a function of $J$.   Substituting the function into the criterion (2.28) we finally obtain a simple relation
 $$0 \le J \le {3\over 8n}\left(1+\left({1\over2}-{3\over 8n^2}\right)B^2\right) +O(B^4).\eqno{(3.21)}$$
which is the stability criterion of the rotating string in the deformed spacetime.  We thus see that magnetic fluxes have inclination to improve the stability of the string solutions.

   Note that, although the metric (3.7), which represents a deformed $AdS_5 \times S^5$, is different form that in our previous paper [16], which represents a deformed $AdS_4 \times S^5$, as the string we considered is fixed on the spatial coordinates in $AdS_5$ and $AdS_4$ respectively, the background considered in the both cases  indeed have the same deformed $S^5$ space. (Some calculation errors of the previous paper are corrected in this paper.)
\\

2. $\nu \gg n$:  In this case, using (3.17) and (3.18) we can find that 
$$cos^2\gamma_0= {J_1\over J_1+2J}.\eqno{(3.22)}$$  
As (3.17) tells us that $\nu$ can be expressed as a function of $J_1$ and $cos^2\gamma_0$,  we can therefore substitute (3.22) into (3.16) to obtain a simple form of the string energy 
$${\cal E} = (2J+J_1) ~ \sqrt{1+ \left({J_1\over 2J+J_1 }\right)^2 B^2}\, ,\eqno{(3.23)}$$
which tells us that there is a positive corrected energy raised from the Melvin magnetic flux which deforms the $S^5$.  To analyze the stability of the above rotating string solution we can follow the previous argument to see that the criterion (2.31), $0 \le sin^2\gamma_0 \le {3\over 4}$, could also be used here.  Now, substituting the relation (3.22) into the criterion (2.31) we finally obtain the criterion of a stable rotating string 
$$ {J\over J_1} \le {3\over2}. \eqno{(3.24)}$$
which shows that the magnetic fluxes does not change  the stability property of the string solutions.  Note that as the correction to the rotating classical string energy is positive then, from the AdS/CFT point of view, the correction of the anomalous dimensions of operators in the dual SYM theory will be positive.

\section{Pulsating Strings in Electric and Magnetic Fields Deformed $AdS_5 \times S^5$}
We will follow the method in [25] to find the possible pulsating string solutions in the electric and magnetic field deformed spacetimes. 
\subsection{Pulsating String Solution in Electric Field Deformed $S^5$}
To proceed, we first identify $t$ with $\tau$ and $\varphi_1$ with $m\sigma$ to allow for multiwrapping and consider a circular pulsating string expanding and contracting on $S^5$. The Nambu-Goto action corresponding to the metric (2.13) becomes
$$S = - m \int dt  \left(1-E^2\right)^{1/4}cos\gamma \,\sqrt{\left(1-E^2\right)^{-1/2}- \dot \gamma^2 - sin^2\gamma\,g_{ij}\dot\phi^i\dot\phi^j}, \eqno{(4.1)}$$
where $g_{ij}$ is the metric on $S_3$ described in (2.13) and $\phi^i$ refers to the coordinates on $S_3$. The canonical momenta calculated from (4.1) are
$$\Pi_\gamma =  {m\left(1-E^2\right)^{1/4}cos\gamma \, \dot\gamma \over \sqrt{\left(1-E^2\right)^{-1/2}- \dot \gamma^2 - sin^2\gamma\, g_{ij}\dot\phi^i\dot\phi^j}}, \eqno{(4.2)}$$
$$\Pi_i = {m\left(1-E^2\right)^{1/4}cos\gamma\, sin^2\gamma \,g_{ij}\dot\phi^j \over \sqrt{\left(1-E^2\right)^{-1/2}- \dot \gamma^2 - sin^2\gamma\, g_{ij}\dot\phi^i\dot\phi^j}}. \eqno{(4.3)}$$
In terms of the canonical momenta the Hamiltonian becomes
$$H^2  = {1\over \sqrt {1-E^2}}\left[\Pi_\gamma^2+ {g^{ij}\Pi_i\Pi_j\over sin^2\gamma}\right]+ m^2 cos^2\gamma. \eqno{(4.4)}$$
In the case of $E=0$ the square of $H$ looks like the Hamiltonian for a particle on $S_5$ with an angular dependent potential.  If we are interested in large
quantum numbers, the potential may be considered as a perturbation and we can proceed by considering free wavefunctions on $S_5$ and then do first order perturbation theory to find the correction.  Denoting the total $S_5$ angular momentum quantum number by $L$ and the total angular momentum quantum number on $S_3$ by $J$ the corresponding zero-order wavefunction had been found in [25].   To the first order correction it was found that 
$$H^2|_{E=0} =  L(L+4) +  m^2\frac{L^2-J^2}{2L^2}, \eqno{(4.5)}$$
which was calculated in eq.(2.9) of [25].  Therefore the spectrum corresponding to the Hamiltonian (4.4) becomes
$$H^2 = {1\over \sqrt {1-E^2}}\, L(L+4) +  m^2\frac{L^2-J^2}{2L^2}. \eqno{(4.6)}$$
As $H^2 > H^2|_{E=0}$ the corrections to the anomalous dimensions of operators in the dual SYM theory are positive.

\subsection{Pulsating String Solution in Magnetic Field Deformed $S^5$}
As before, we identify $t$ with $\tau$ and $\varphi_1$ with $m\sigma$ to allow for multiwrapping and consider a circular pulsating string expanding and contracting on $S^5$.   The Nambu-Goto action corresponding to the metric (3.8) becomes
$$S = -m \int dt \,cos\gamma\, \sqrt{1- \dot \gamma^2 - sin^2\gamma \,g_{ij}\dot\phi^i\dot\phi^j}, \eqno{(4.7)}$$
where $g_{ij}$ is the metric on $S_3$ described in (3.8) and $\phi^i$ refers to the coordinates on $S_3$.   As the action is independent of magnetic field $B$ the pulsating string solution is exactly like that in [25].

   It shall be remarked that this trivial behavior may be traced to the special pulsating string we assumed.  Other pulsating string solutions like those described [26] are difficult to be investigated and are expected to have nontrivial behaviors.  Note also that we have only considered the strings on the deformed $S^5$ spacetimes in this paper.  The coordinates of the deformed $AdS_5$ spacetime in (2.12) or (3.7) are mathematically complex and it is difficult to study the strings therein. The problems remain to be investigated.


\section{Conclusion}
In this paper we apply the transformation of mixing azimuthal and internal coordinate or  mixing time and internal coordinate to the 11D M-theory with a stack N M2-branes to find  the spacetime of a stack of  N D2-branes with magnetic or electric flux  in 10 D IIA string theory, after the Kaluza-Klein reduction.  We perform the T duality to the spacetimes to find the backgrounds of a stack of  N D3-branes with magnetic or electric flux.  In the near-horizon limit the background becomes the magnetic or electric field deformed $AdS_5 \times S^5$.   In contrast to the previous study in which the background were magnetic-flux deformed $AdS_5 \times S^4$ and $AdS_4 \times S^5$, the isometry groups of  $AdS_5 \times S^5$ is $SO(2,4) \times SO(6)$ which are exactly the conformal group and R-symmetry of $N=4$ supper Yang-Mills theory, the backgrounds studied in this paper will be more phenomenally interesting.  

   We have found the classical spinning string solutions which are rotating in the deformed $S^5$ with three angular momenta $J$ in the three rotation planes.  The relations between the classical string energy and its angular momenta are obtained and results show that the external magnetic and electric fluxes will increase the string energy.  Therefore, from the AdS/CFT point of view,  the corrections to the anomalous dimensions of operators in the dual SYM theory will be positive.  We have investigated the small fluctuations in these solutions and discuss the effects of magnetic and electric fields on the  stability of these classical rotating string solutions.  We have also found the possible solutions of string pulsating on the deformed spacetime.  From the corrections to the pulsating string energy we see that the corrections to the anomalous dimensions of operators in the dual SYM theory are non-negative.

  As the supersymmetry is broken by the magnetic or electric field there will  in general  appear tachyon when the string is in the magnetic or electric deformed spacetime.  However, as the classical solutions studied in this paper correspond to the states with large quantum number the tachyon will not be shown.   It will be interesting to find the string spectrum by following the method in [27] to see more properties of the effects of the magnetic or electric field on the rotating and pulsating string solutions.  Also, the correspondence between spin chain [28] and classical strings  in the electric or magnetic filed  deformed $AdS^5 \times S^5$, which  can improve our understanding of the AdS/CFT correspondence, are deserved to be investigated.

   Finally, it is known from $AdS_5 \times S^5$ case that in order to establish the correspondence  relation between the energy and spins to the anomalous dimension  it is important to prove that the non Cartan elements of the angular momentum are vanishing.  Also, as the energy of a string state is conjectured to be equal to the scaling dimension of the dual gauge theory operator it is important to find the Yang-Mills operators corresponding to the classical rotating and pulsating strings in the magnetic or electric field deformed spacetimes. (The corresponding operators in the undeformed backgrounds had been studied in [4,5,6,25].)  To complete the analysis from AdS/CFT point of view we shall also develop Bethe ansatz analysis [29,30] and compare the above result to that in SYM side.  These important problems are left to further research.

\vspace{2cm}

{\Large \bf  References}
\begin{enumerate}
{\small
\item J.~M.~Maldacena, ``The large N limit of superconformal field theories and supergravity,'' Adv.\ Theor.\ Math.\ Phys.\  {\bf 2}, 231 (1998) [Int.\ J.\ Theor.\ Phys.\  38 (1999) 1113  [hep-th/9711200]; S.~S.~Gubser, I.~R.~Klebanov and A.~M.~Polyakov, ``Gauge theory correlators from non-critical string theory,'' Phys.\ Lett.\ B428 (1998) 105 [hep-th/9802109]; E.~Witten, ``Anti-de Sitter space and holography,'' Adv.\ Theor.\ Math.\ Phys.\   2 (1998) 253 [hep-th/9802150].
\item O.~Aharony, S.~S.~Gubser, J.~M.~Maldacena, H.~Ooguri and Y.~Oz, ``Large N field theories, string theory and gravity,'' Phys.\ Rept. 323 (2000) 183 (2000) [hep-th/9905111]; E.~D'Hoker and D.~Z.~Freedman, ``Supersymmetric gauge theories and the AdS/CFT correspondence,''  [hep-th/0201253].
\item D.~Berenstein, J.~M.~Maldacena and H.~Nastase, ``Strings in flat
space and pp waves from N = 4 super Yang Mills'',  JHEP  0204 (2002) 013 [hep-th/0202021].
\item  S.~S.~Gubser, I.~R.~Klebanov and A.~M.~Polyakov, ``A semi-classical limit of the gauge/string correspondence,'' Nucl.\ Phys.\ B636 (2002) 99 [hep-th/0204051]. 
\item S.~Frolov and A.~A.~Tseytlin, ``Multi-spin string solutions in
$AdS_5 \times S^5$,'' Nucl.\ Phys.\ B668 (2003) 77 [hep-th/0304255].
\item S.~Frolov and A.~A.~Tseytlin, ``Quantizing three-spin string
solution in $AdS_5 \times S^5$,'' JHEP 0307 (2003) 016 [hep-th/0306130].
\item S.~Frolov and A.~A.~Tseytlin, ``Rotating string solutions: AdS/CFT duality in non-supersymmetric sectors,'' Phys.\ Lett.\ B570 (2003) 96 [hep-th/0306143]; S.A. Frolov, I.Y. Park, A.A. Tseytlin, ``On one-loop correction to energy of spinning strings in $S^5$,''Phys.Rev. D71 (2005) 026006 [hep-th/0408187]; I.Y. Park, A. Tirziu, A.A. Tseytlin, ``Spinning strings in $AdS_5 \times  S^5$: one-loop  correction to energy in SL(2) sector,'' JHEP 0503 (2005) 013 [hep-th/0501203]; N. Beisert, A.A. Tseytlin, K. Zarembo, ``Matching quantum strings to quantum spins: one-loop vs. finite-size corrections,''Nucl.Phys. B715 (2005) 190-210 [hep-th/0502173].
\item G.~Arutyunov, S.~Frolov, J.~Russo and A.~A.~Tseytlin, ``Spinning strings in $AdS_5 \times S^5$ and integrable systems,'' Nucl.\ Phys.\ B671 (2003) 3 [hep-th/0307191]; G.~Arutyunov, J.~Russo and A.~A.~Tseytlin, ``Spinning strings in $AdS_5 \times S^5$: New integrable system relations,'' Phys.\ Rev.\ D69 (2004) 086009 [hep-th/0311004].
\item A.~A.~Tseytlin, ``Spinning strings and AdS/CFT duality,'' [hep-th/0311139]; J, Plefka, ``Spinning strings and integrable spin chains in the AdS/CFT correspondence,'' [hep-th/0507136].
\item  O.~Lunin and J.~Maldacena, ``Deforming field theories with U(1) x U(1) global symmetry and their gravity duals,'' JHEP  0505  (2005)  033  [hep-th/0502086]. 
\item R.~G.~Leigh and M.~J.~Strassler, ``Exactly marginal operators and duality in four-dimensional N=1 supersymmetric gauge theory,'' Nucl.\ Phys.\ B447 (1995) 95 [hep-th/9503121].
\item S.~A.~Frolov, R.~Roiban and A.~A.~Tseytlin, ``Gauge-string duality for superconformal deformations of N = 4 super Yang-Mills theory,'' JHEP 0507 (2005) 045 [hep-/0503192];  S.  Frolov, ``Lax Pair for Strings in Lunin-Maldacena Background,'' JHEP 0505 (2005) 069 [ hep-th/0503201]; N.P. Bobev, H. Dimov, R.C. Rashkov, ``Semiclassical strings in Lunin-Maldacena background,'' [hep-th/0506063]; R. de M. Koch, N. Ives, J. Smolic and M. Smolic, ``Unstable Giants,'' [hep-th/0509007].
\item  S.M. Kuzenko, A.A. Tseytlin, ``Effective action of beta-deformed N=4 SYM theory and AdS/CFT,''  [hep-th/0508098]; J. G. Russo, ``String spectrum of curved string backgrounds obtained by T-duality and shifts of polar angles,'' JHEP 0509 (2005) 031 [hep-th/0508125]; M. Spradlin, T. Takayanagi and A. Volovich, ``String Theory in Beta Deformed Spacetimes,'' [hep-th/0509036];  R. C. Rashkov, K. S. Viswanathan and Y. Yang, ``Generalizations of Lunin-Maldacena transformation on the $AdS_ 5 \times S^5$ background,'' [hep-th/0509058];  S. Ryang,``Rotating Strings with Two Unequal Spins in Lunin-Maldacena Background,''  [hep-th/0509195]; H. Ebrahim and A. E. Mosaffa, ``Semiclassical String Solutions on 1/2 BPS Geometries,'' JHEP 0501 (2005) 050 [hep-th/0501072]; H. Ebrahim, ``Semiclassical Strings Probing NS5 Brane Wrapped on $S^5$,'' [hep-th/0511228].
\item I.R. Klebanov, A.A. Tseytlin, ``Gravity Duals of Supersymmetric SU(N) x SU(N+M) Gauge Theories,'' Nucl.Phys. B578 (2000) 123-138 [hep-th/0002159]; M.  Schvellinger, ``Spinning and rotating strings for N=1 SYM theory and brane constructions,'' JHEP 0402 (2004) 066 [hep-th/0309161]; Xiao-Jun Wang, ``Spinning Strings on Deformed $AdS_5 \times T^{1,1}$ with NS B-field,'' Phys. Rev. D72 (2005) 086006 [hep-th/0501029].
\item U. Gursoy and C. Nunez, ``Dipole Deformations of N=1 SYM and Supergravity backgrounds with U(1) X U(1) global symmetry,''  Nucl.Phys. B725 (2005) 45-92 [hep-th/0505100]; N.P. Bobev, H. Dimov, R.C. Rashkov, ``Semiclassical Strings, Dipole Deformations of N=1 SYM and Decoupling of KK Modes,'' [hep-th/0511216]; S. S. Pal, ``$\beta$-deformations, potential and KK modes,'' Phys. Rev. D72 (2005) 065006 [hep-th/0505257]; R. C. Rashkov, K. S. Viswanathan and Yi Yang, ``Semiclassical Analysis of String/Gauge Duality on Non-commutative Space, ''Phys.Rev. D70 (2004) 086008 [hep-th/0404122
].
\item Wung-Hong Huang, `` Multi-spin String Solutions in Magnetic-flux Deformed $AdS_n \times S^m$ Spacetime, '' JHEP 0512 (2005) 013 [hep-th/0510136].
\item F. Dowker, J. P. Gauntlett, D. A. Kastor and Jennie Traschen, ``Pair Creation of Dilaton Black Holes,'' Phys.Rev. D49 (1994) 2909-2917  [hep-th/9309075].
\item F.~Dowker, J.~P.~Gauntlett, D.~A.~Kastor and J.~Traschen, ``The decay of magnetic fields in Kaluza-Klein theory,'' Phys.\ Rev.\ D52 (1995) 6929 [hep-th/9507143]; M.~S.~Costa and M.~Gutperle, ``The Kaluza-Klein Melvin solution in M-theory,'' JHEP 0103 (2001) 027 [hep-th/0012072].
\item M.A. Melvin, ``Pure magnetic and electric geons,'' Phys. Lett. 8 (1964) 65; 
\item G.~W.~Gibbons and D.~L.~Wiltshire, ``Space-time as a membrane in higher dimensions,'' Nucl.\ Phys.\ B287 (1987) 717 [hep-th/0109093]; G.~W.~Gibbons and K.~Maeda, ``Black holes and membranes in higher dimensional theories with dilaton fields,'' Nucl.\ Phys.\ B298 (1988) 741; A. Chodos and S. Detweiler, Gen. Rel. and Grav. 14 (1982) 879; G.W. Gibbons and D.L. Wiltshire, Ann.Phys. 167 (1986) 201; erratum  ibid. 176 (1987) 393.
\item T. Friedmann and H. Verlinde, ``Schwinger pair creation of Kaluza-Klein particles: Pair creation without tunneling,'' Phys.Rev. D71 (2005) 064018 [hep-th/0212163]; L. Cornalba and M.S. Costa, ``A New Cosmological Scenario in String Theory,'' Phys.Rev. D66 (2002) 066001 [hep-th/0203031]; L. Cornalba, M.S. Costa and C. Kounnas, ``A Resolution of the Cosmological Singularity with Orientifolds,'' Nucl.Phys. B637 (2002) 378-394 [hep-th/0204261].
\item C.~G.~Callan, J.~A.~Harvey and A.~Strominger, ``Supersymmetric string solitons,'' [hep-th/9112030]; A.~Dabholkar, G.~W.~Gibbons, J.~A.~Harvey and F.~Ruiz Ruiz, ``Superstrings And Solitons,'' Nucl.\ Phys.\ B  340 (1990) 33; G.T. Horowitz and A.~Strominger, ``Black strings and P-branes,'' Nucl.\ Phys.\ B  360 (1991) 197.
\item P. Ginsparg and C. Vafa, Nucl. Phys. B289 (1987) 414; T. Buscher, Phys. Lett. B159 (1985) 127; B194 (1987) 59; B201 (1988) 466.
\item  S. F. Hassan, ``T-Duality, Space-time Spinors and R-R Fields in Curved Backgrounds,'' Nucl.Phys. B568 (2000) 145 [hep-th/9907152].
\item  J. Engquist, J. A. Minahan and K. Zarembo, ``Yang-Mills Duals for Semiclassical Strings,'' JHEP 0311 (2003) 063 [hep-th/0310188].
\item J.~A.~Minahan, ``Circular semiclassical string solutions on $AdS_5 \times S^5$,'' Nucl.\ Phys.\ B648 (2003) 203 [hep-th/0209047]; M. Alishahiha, A. E. Mosaffa and H. Yavartanoo, ``Multi-spin string solutions in AdS Black Hole and confining backgrounds,''  Nucl.Phys. B686 (2004) 53 [hep-th/0402007];  A. Khan and A. L. Larsen, ``Spinning Pulsating String Solitons in $AdS_5 \times S^5$,'' Phys. Rev. D69 (2004) 026001[hep-th/0310019]; A. Khan and A. L. Larsen, ``Improved Stabality for Pulsating Multi-spin String Solitons,'' [hep-th/0502063].
\item  J.~G.~Russo and A.~A.~Tseytlin, ``Exactly solvable string models of curved space-time backgrounds,'' Nucl.\ Phys.\ B449 (1995) 91 [hep-th/9502038]; ``Magnetic flux tube models in superstring theory,'' Nucl.\ Phys.\ B461 (1996) 131 [hep-th/9508068].
\item M.~Kruczenski, ``Spin chains and string theory,'' Phys. Rev. Lett. 93 (2004) 161602. h[ep-th/0311203]; R.~Hernandez and E.~Lopez, ``The SU(3) spin chain sigma model and string theory,'' JHEP 0404 (2004) 052 [hep-th/0403139]; S. Bellucci, P. Y. Caesteill, J. F. Morales and C. Sochichi, ``SL(2) spin chain and spinning strings on $AdS_5 \times S^5$,'' Nucl.\ Phys.\ B707 (2005) 303 [hep-th/0409086];  B. Stefanski, jr. and  A.A. Tseytlin, ``Large spin limits of AdS/CFT and generalized Landau-Lifshitz equations,'' JHEP 0405 (2004) 042 [hep-th/0404133]; ``Super spin chain coherent state actions and $AdS_5 \times S^5$ superstring,''  Nucl.Phys. B718 (2005) 83 [hep-th/0503185]; A.A. Tseytlin,  ``Semiclassical strings and AdS/CFT,'' [hep-th/0409296]; S. Ryang, ``Circular and Folded Multi-Spin Strings in Spin Chain Sigma Models,''  JHEP 0410 (2004) 059 [hep-th/0409217].
\item J. A. Minahan and K. Zarembo, ``The Bethe-Ansatz for N=4 Super Yang-Mills,'' JHEP 0303 (2003) 013 [hep-th/0212208]; V.A.Kazakov, A.Marshakov, J.A.Minahan and K.Zarembo, ``Classical/quantum integrability in AdS/CFT,'' JHEP 0405 (2004) 024 [hep-th/0402207].
\item  K. Peeters, J. Plefka, and M. Zamaklar, ``Splitting spinning strings in AdS/CFT,'' [hep-th/0410275]; "Splitting strings and chains,'' [hep-th/0501165]; S. Schafer-Nameki, M. Zamaklar and K. Zarembo, ``Quantum corrections to spinning strings in AdS(5) x S(5) and Bethe ansatz: a comparative study,'' JHEP 0509 (2005) 051 [hep-th/0507189];  S. Schafer-Nameki and M. Zamaklar, ``Stringy sums and corrections to the quantum string Bethe ansatz,''  JHEP 0510 (2005) 044 [hep-th/0509096]; N. Beisert and A. A. Tseytlin, ``On Quantum Corrections to Spinning Strings and Bethe Equations,'' Phys.Lett. B629 (2005) 102 [hep-th/0509084]; G. Arutyunov and M. Zamaklar, ``Linking Backlund and Monodromy Charges for Strings on $AdS_5 x S^5$,'' [hep-th/0504144].
}
\end{enumerate}
\end{document}